\documentclass{IEEEojcsys}

\usepackage[colorlinks,urlcolor=blue,linkcolor=blue,citecolor=blue]{hyperref}

\usepackage{color,array}

\usepackage{graphicx}
\usepackage{amsmath}
\usepackage{amssymb}
\usepackage{cite}
\usepackage{amsmath,amssymb,amsfonts}
\usepackage{graphicx}
\usepackage{textcomp}
\usepackage{algorithmic}
\usepackage{hyperref}

\usepackage{xcolor}
\usepackage{listings}

\usepackage{xparse}

\usepackage{subcaption}

\NewDocumentCommand{\codeword}{v}{%
\texttt{\textcolor{black}{#1}}%
}

\usepackage[ruled,vlined, linesnumbered]{algorithm2e}

\jvol{00}
\jnum{XX}
\paper{1234567}
\pubyear{2021}
\receiveddate{XX September 2021}
\accepteddate{XX October 2021}
\publisheddate{XX November 2021}
\currentdate{XX November 2021}
\doiinfo{OJCSYS.2021.Doi Number}

\begin{document}

\sptitle{Article Category}

\title{Inside madupite: \\ Technical Design and Performance}

\editor{This paper was recommended by Associate Editor F. A. Author.}

\author{M. Gargiani\affilmark{1}}

\author{R.~Sieber\affilmark{2}}
\author{P.~Pawlowsky\affilmark{2}}

\author{J. Lygeros\affilmark{1}  (Member, IEEE)}

\affil{Automatic Control Laboratory (IfA), Physikstrasse 3,
8092 Zurich, Switzerland} 
\affil{ETH Zurich, Rämistrasse 101, 8092 Zürich} 

\corresp{CORRESPONDING AUTHOR: M. Gargiani (e-mail: \href{mailto:gmatilde@ethz.ch}{gmatilde@ethz.ch})}
\authornote{This work was supported by the European Research Council under the Horizon 2020 Advanced under Grant 787845 (OCAL).}

\markboth{PREPARATION OF PAPERS FOR IEEE OPEN JOURNAL OF CONTROL SYSTEMS}{M. Gargiani {\itshape ET AL}.}

\begin{abstract}
In this work, we introduce and benchmark madupite, a newly proposed high-performance solver designed for large-scale discounted infinite-horizon Markov decision processes with finite state and action spaces. After a brief overview of the class of mathematical optimization methods on which madupite relies, we provide details on implementation choices, technical design and deployment. We then demonstrate its scalability and efficiency by showcasing its performance on the solution of Markov decision processes arising from different application areas, including epidemiology and classical control.
Madupite sets a new standard as, to the best of our knowledge, it is the only solver capable of efficiently computing exact solutions for large-scale Markov decision processes, even when these exceed the memory capacity of modern laptops and operate in near-undiscounted settings. This is possible as madupite can work in a fully distributed manner and therefore leverage the memory storage and computation capabilities of modern high-performance computing clusters. This key feature enables the solver to efficiently handle problems of medium to large size in an exact manner instead of necessarily resorting to function approximations. Moreover, madupite is unique in allowing users to customize the solution algorithm to better exploit the specific structure of their problem, significantly accelerating convergence especially in large-discount factor settings.
Overall, madupite represents a significant advancement, offering unmatched scalability and flexibility in solving large-scale Markov decision processes.
\end{abstract}


\maketitle

\section{INTRODUCTION}

Markov Decision Processes (MDPs) offer a powerful and general mathematical framework for modeling sequential decision-making problems across a wide range of fields, including epidemiology~\cite{sis_model}, finance~\cite{finance_MDP}, robotics~\cite{robotics_mdp} and agriculture~\cite{agriculture_mdp}. Dynamic programming, introduced by Bellman in the 1950s, remains a foundational method for solving MDPs~\cite{Bellman:1957}. However, practical applications of dynamic programming algorithms face significant challenges, as they typically suffer from slow convergence in high discount factor regimes or limited scalability in large-scale settings~\cite{DB_book}. Both aspects are essential for the exact solution of MDPs derived from real-world applications, which are often modeled with very high discount factors~\cite{lecture_gros} and involve extremely large state and action spaces, a challenge commonly referred to as \textit{curse of dimensionality}~\cite{DB_book}. To have any realistic chance of solving such problems exactly, it is crucial to address both of these dimensions.

Recent advances in algorithmic design have introduced inexact policy iteration (iPI) methods, a new class of dynamic programming techniques that overcome some of the traditional limitations by achieving rapid convergence, even in scenarios with discount factors approaching one~\cite{gargiani_2023, gargiani_iPI}. Nevertheless, solving large-scale MDPs requires more than algorithmic advances, it demands leveraging high-performance computing (HPC) platforms. Distributed and parallel implementations are crucial for harnessing the full computational power of modern HPC clusters. In shared-memory environments, parallelization accelerates computation by dividing tasks across multiple cores, enabling faster execution through concurrent processing. In distributed-memory settings, these implementations go further by distributing both data and computation across multiple nodes, overcoming memory limitations and making it possible to solve MDPs that are otherwise infeasible on a single machine, all while maintaining low communication overhead.

While reinforcement learning has made significant strides in tackling large MDPs through function approximation~\cite{Sutton_RLbook}, such approaches introduce added complexity to the optimization process and make it difficult to assess how close the resulting policy is to the true optimal solution~\cite{rl_hard}. This lack of quantifiable performance guarantees can be a major drawback in applications requiring solutions with guaranteed near-optimality. In contrast, exact dynamic programming methods offer theoretical guarantees but face scalability challenges. Existing toolkits, such as pymdptoolbox~\cite{pymdptoolbox} and mdpsolver~\cite{MDPsolver}, illustrate these issues. Although the latter supports multi-core execution, it is limited to shared-memory parallelism and thus constrained by the memory and compute capacity of a single machine. This makes it unsuitable for deployment on an HPC cluster. Furthermore, these toolkits leverage optimistic policy iteration (OPI) as solution method, which results in significantly slower convergence in high discount factor regimes, which are particularly relevant in many real-world applications~\cite{gargiani_iPI, gargiani_2023, lecture_gros}.

To address these limitations, we introduce madupite, a solver that combines algorithmic rigor with technical innovation. Madupite leverages inexact policy iteration methods and distributed HPC infrastructure to efficiently solve infinite-horizon discounted MDPs with finite state and action spaces, scaling beyond what a single laptop can handle. Our solver accelerates computation through parallelism while maintaining efficiency in high discount factor regimes, thanks to the convergence guarantees and contraction properties of iPI algorithms. In addition, madupite is purpose-built for distributed-memory environments, enabling it to handle problems that exceed the memory and processing capacity of individual nodes. Finally, the solver offers extensive customization, allowing users to tailor solution strategies to the structural characteristics of their MDPs. As illustrated in details in its documentation~\cite{madupite_doc}, users can indeed configure the linear solver, stopping condition, preconditioner matrix and other many other parameters, tailoring the solution method to their problem in order to enhance convergence. 

Implemented in modern C++20 and equipped with an intuitive Python API, madupite is engineered to fully utilize HPC cluster resources. Built on top of the PETSc (Portable, Extensible Toolkit for Scientific Computation) scientific computing toolkit~\cite{petsc-efficient, petsc-user-ref, petsc-web-page}, our solver supports both fully distributed and shared-memory parallel execution via MPI—enabling the efficient solution of MDPs with millions of states.

While a previous publication in the Journal of Open Source Software (JOSS) primarily focused on the codebase and general purpose of madupite~\cite{gargiani_madupite}, this paper focuses on the solver’s technical design, architectural decisions, and performance analysis. We provide a detailed account of the core algorithms, software architecture, and engineering practices that underpin madupite's efficiency, as well as its integration with PETSc.
Then, through extensive benchmarks and real-world-inspired case studies, including infectious disease modeling and classical control applications, we showcase madupite’s scalability, flexibility, and superior performance. Our results highlight its capability to efficiently and exactly solve large-scale MDPs, promoting reproducible research and supporting future developments in scalable exact solution methods for MDPs. 

The remainder of the paper is organized as follows. Section~\ref{sec: problem_setting} introduces the formal problem setting of infinite-horizon discounted MDPs with finite state and action spaces, including both intuitive and mathematical formulations.
Section~\ref{sec: madupite} presents madupite solver in detail. In particular, in Subsection~\ref{sec: madupite}-\ref{subsec: methods}, we describe iPI methods that form the algorithmic foundation of the solver. Subsection~\ref{sec: madupite}-\ref{subsec: implementation_choices} outlines key implementation choices, including distributed memory parallelism (\ref{sec: madupite}-\ref{subsec: implementation_choices}.\ref{subsubsec: distributed_mem_par}), data structures and layouts (\ref{sec: madupite}-\ref{subsec: implementation_choices}.\ref{subsubsec: datastructures_layouts}), and details on the implementation of iPI methods (\ref{sec: madupite}-\ref{subsec: implementation_choices}.\ref{subsubsec: implementation_details_iPI}). Subsection~\ref{sec: madupite}-\ref{subsec: pythonAPI_deployment} introduces the Python API and provides guidance for deploying madupite in practical settings.
Section~\ref{sec: performance_evaluation} evaluates the performance of madupite. Subsection~\ref{sec: performance_evaluation}-\ref{subsec: performance_evaluation}.\ref{subsubsec:amdahlslaw} analyzes parallel efficiency using Amdahl’s law. Subsection~\ref{sec: performance_evaluation}-\ref{subsec: performance_evaluation}.\ref{subsubsec:solvercustomization} explores the impact of solver customization on convergence. Subsection \ref{sec: performance_evaluation}-\ref{subsec: performance_evaluation}.\ref{subsubsec:memorylimitations} demonstrates madupite’s ability to solve MDPs that exceed laptop memory constraints by leveraging HPC clusters. Subsection \ref{sec: performance_evaluation}-\ref{subsec: performance_evaluation}.\ref{subsubsec:comparisonSOTA} compares madupite with existing state-of-the-art MDP solvers. This is followed by case studies: Subsection \ref{sec: performance_evaluation}-\ref{subsec:infectiondisease} addresses infectious disease modeling, Subsection \ref{sec: performance_evaluation}-\ref{subsec:pendulum} covers the classic inverted pendulum control problem, and Subsection \ref{sec: performance_evaluation}-\ref{subsec: maze} examines a deterministic 2D maze environment.
Finally, Section~\ref{sec: conclusions} concludes the paper, summarizing key contributions and outlining directions for future work.






\section{Problem Setting}\label{sec: problem_setting}

Madupite solves discounted infinite-horizon MDPs with finite state and action spaces. In broad terms, MDPs provide a general mathematical framework for modeling memoryless stochastic sequential decision processes (MSSDPs) arising in diverse fields~\cite{puterman_mdp}. In this section, we present the core mathematical concepts that define these models, offering a foundation for transitioning from practical applications to their abstract mathematical formulations. Starting from a MSSDP, by identifying and interpreting the key components, we can employ madupite to create its MDP representation and efficiently compute the solution, which can then be applied to the original system to ensure optimal performance.

Consider a complex system, for instance a robot navigating in a maze or the spread of a viral disease in a population. The state of our system changes over time based on two factors: the actions we choose to apply to it and the random noise that affects it, which we can not actively control. The goal is to select the best action at each step, where the optimality of a given action is determined by a specific metric that we aim at minimizing.
To explain further and in general terms, each time we take an action in a given state, there’s a cost, referred to as the stage-cost. We are interested in the expected value of the total sum of these costs over time. However, we don't treat all costs equally: we apply a multiplicative discount factor, meaning we give more weight to costs that occur sooner rather than those that happen later in the horizon. This makes immediate costs more significant in our decision-making.
In summary, the goal is to choose actions that minimize the expected sum of these discounted stage-costs over the entire time period, where the expectation is the metric considered to account for the uncertainty caused by noise in the system.

Now if we translate these concepts mathematically, we can use a 5-elements tuple $\left\{\mathcal{S},\,\mathcal{A},\,P,\,g,\, \gamma \right\}$, where, without loss of generality, $\mathcal{S} = \left\{ 1,\dots, n\right\}$ and $\mathcal{A} = \left\{1,\dots,m\right\}$ are the state and action spaces, respectively; $P:\mathcal{S}\times\mathcal{S}\times \mathcal{A}\rightarrow [0,1]$ is the transition probability function, where $P(s,\,s',\,a)$ is the probability of transitioning to state $s'$ when we play action $a$ in state $s$ and $\sum_{s'\in\mathcal{S}} P(s,\,s',\,a) = 1$ for all $s\in\mathcal{S}$ and $a\in \mathcal{A}$; $g:\mathcal{S}\times\mathcal{A}\rightarrow \mathbb{R}$ is the stage-cost function, where $g(s,a)$ is the cost associated to the $(s,a)$ state-action pair; and $\gamma\in(0,1)$ is the discount factor, where higher values are associated to less myopic optimal action-strategies and viceversa.
Notice that the transitions are not necessarily deterministic since, as discussed, the evolution of the system may also be affected by noise. The goal is to compute the infinite sequence of actions that minimizes the following quantity for all $s\in \mathcal{S}$
\begin{equation}\label{eq:V_pi}
     \lim_{T\rightarrow\infty}\mathbb{E}\left[ \sum_{t=0}^{T-1} \gamma^t g(s_t, \,a_t)\,\middle | s_0 = s \right]\,,
\end{equation}
where the expected value is taken with respect to the corresponding probability measure over the space of sequences. 
Instead of considering sequences of actions, we can equivalently define a function $\pi:\mathcal{S}\rightarrow \mathcal{A}$, called policy, where $\pi(s)$ is the action to play in state $s$. We denote with $V_{\pi}(s)$ the quantity introduced in Eq.~\eqref{eq:V_pi} and we call optimal and denote it with $\pi^*$ any policy $\pi$ which attains the minimum of~\eqref{eq:V_pi}. Finding the optimal sequence of actions therefore corresponds to finding an optimal policy. Given the finiteness of the state and action spaces, there only exists a finite set of policies $\Pi$ with dimension at most $m^n$. Clearly, due to the exponential size of this space, designing efficient methods with strong improvement-per-iteration properties becomes crucial.

Because of the finiteness of the spaces, these quantities can also be expressed in vector and matrix notation, which allows one to re-write the problem more compactly and also makes it easier to transition to madupite. In particular, $V_{\pi}\in \mathbb{R}^n$ is the vector whose $s$-th entry corresponds to~\eqref{eq:V_pi}; $P\in\mathbb{R}^{n\times n \times m}$ is the 3-dimensional tensor where the $a$-th slice is an $n\times n$ row-stochastic matrix whose $(s, s')$-element corresponds to $P(s, s', a)$; and finally $g \in \mathbb{R}^{n\times m}$ is the matrix whose $(s,a)$-th element corresponds to $g(s,a)$. We also denote with $g_{\pi}$ the $n$-dimensional vector whose $s$-the entry is $g(s,\pi(s))$, and with $P_{\pi}$ the $n\times n$ row-stochastic matrix whose $(s, s')$-entry is $P(s,\,s',\,\pi(s))$. By deploying this notation, we can compactly express the cost associated to a policy $\pi$ as the solution of the following linear system of equations
\begin{equation}\label{eq:bellman_eq_pi}
    V_{\pi} = g_{\pi} + \gamma P_{\pi}V_{\pi}\,,
\end{equation}
while the cost associated to the optimal policy as the solution of the following system of non-linear equations
\begin{equation}\label{eq:bellman_eq_star}
    V_{\pi^*} = \min_{\pi \in \Pi} \left\{ g_{\pi} + \gamma P_{\pi}V_{\pi^*} \right\}\,.
\end{equation}
These equations are known as Bellman's equations and their fundamental properties are at the foundations of classical solutions methods~\cite{DB_book}. They do also provide a way to evaluate the ``optimality'' of a cost-vector $V\in \mathbb{R}^n$. It is indeed possible to evaluate the latter by computing the norm of the residual $$r(V) = V - TV =  V - \min_{\pi\in\Pi}\left\{ g_{\pi} + \gamma P_{\pi}V \right\}\,.$$ Analogously, we can evaluate how ``far-off'' with respect to $V_{\pi}$ is a vector $V\in \mathbb{R}^n$ by computing the norm of $$r_{\pi}(V) = V- T_{\pi}V = V - ( g_{\pi} + \gamma P_{\pi}V )\,.$$ These quantities tell us how well a vector $V\in\mathbb{R}^n$ approximates $V_{\pi^*}$ or $V_{\pi}$, respectively.

Finally, given any $V\in\mathbb{R}^n$, we call greedy any policy $\pi\in \Pi$ such that
\begin{equation}\label{eq:greedy_policy}
    \pi \in \arg\min_{\pi\in \Pi} \left\{ g_{\pi} + \gamma P_{\pi}V\right\}\,.
\end{equation}

\section{MADUPITE}\label{sec: madupite}
This section is dedicated to a description of madupite solver, from the methodologies on which it relies to efficiently tackle MDPs on an algorithmic viewpoint, to more technical implementation details which are key to understand the high-performance characteristics of our solver. Finally, we also provide a general description of its Python interface, which, in combination with the detailed documentation of madupite, allows for an easy deployment of our solver also by non-expert users. 

Madupite is open-source and available at \url{https://github.com/madupite/madupite}. In addition, a detailed documentation with step-by-step instructions on the installation as well as examples and tutorials is available at \url{https://madupite.github.io/}.
\subsection{Methods}\label{subsec: methods}
Madupite solver relies on inexact policy iteration methods to efficiently solve MDPs. It is out of the scope of this paper to review the mathematical details and convergence properties of this class of methods, but we refer the interested readers to~\cite{gargiani_iPI}. We will limit the discussion here to those aspects which are relevant for a correct and informed deployment of madupite solver.

As the name suggests, this class of methods is based on an inexact variant of policy iteration~\cite{DB_book, gargiani_iPI}. As in the classical policy iteration scheme, the methods in this class alternate the extraction of a greedy policy and the computation of the cost associated to it. Differently from the classical policy iteration scheme, the latter step, known as policy evaluation, is solved only inexactly and the approximation level is regulated via a parametric stopping condition. Algorithm~\ref{alg:iPI} offers the pseudocode description of a general inexact policy iteration method, where step 3 corresponds to the greedy policy extraction step, and steps 4-12 correspond to the inexact policy evaluation step. We use $\text{GreedyPolicy}(V)$ to denote the operator that extracts a greedy policy associated to the cost-vector $V$ as described in~\eqref{eq:greedy_policy}, $\text{Preconditioner}(\text{``type''},A)$ for the operator that returns the ``type''-preconditioner for the matrix $A$, and $\text{IterativeSolver}(A, b, \tilde{x})$ to denote the operator that applies one iteration with a well defined iterative solver for linear systems with non-singular coefficient matrices to the linear system $Ax = b$ with $A$ being non-singular and starting from $\tilde{x}$. Different iterative methods can be deployed for that, such as GMRES~\cite{gmres}, Richardson's method~\cite{saad}, SOR~\cite{saad}, TFQRM~\cite{tfqmr} etc. It has been proven both theoretically and empirically that the optimal choice for the inner solver is linked to the structural characteristics of the underlying MDPs~\cite{gargiani_iPI}.  
\begin{algorithm}
\caption{\\A General Inexact Policy Iteration Method}\label{alg:iPI}
\SetKwInOut{Input}{Input}
\SetKwInOut{Output}{Output}
\Input{$V_{0} \in \mathbb{R}^n$, $\textit{tol}>0$, $\alpha \in (0,1)$, $N_o, \,N_i \in \mathbb{Z}_{>0}$, ``type''}
\Output{$V_{k}$ for some $k$}
$k \leftarrow 0$\;
\While {$\Vert r(V_{k})\Vert_{\infty} > \textit{tol}$}{
    $\pi_{k+1} \leftarrow \text{GreedyPolicy}(V_{k})$\;
    $J_{\pi_{k+1}} \leftarrow I - \gamma P_{\pi_{k+1}}$\;
    $D_{k} \leftarrow \text{Preconditioner}($\text{``type''}$, J_{\pi_{k+1}})$;\\
    $\theta_{0} \leftarrow V_k$\;
    $i \leftarrow 0$\;
    \While{$\Vert r_{\pi_{k+1}}(\theta_i) \Vert_{\infty}\geq \alpha\Vert r_{\pi_{k+1}}(V_k) \Vert_{\infty}$}{
        ${\theta}_{i+1} \leftarrow \text{IterativeSolver}(D_{k}J_{\pi_{k+1}}, D_{k}g_{\pi_{k+1}}, {\theta}_{i})$\;
        $i \leftarrow i + 1$\;
        \If {$i> N_i$} {break;}
    }
    $V_{k+1} \leftarrow \theta_i$\;
    $k \leftarrow k + 1$\;
    \If {$k> N_o$} {break;}
}
\Return $V_{k},\,\pi_k$
\end{algorithm}
\subsection{Implementation Choices}\label{subsec: implementation_choices}
The core of madupite is written in C++ and leverages PETSc~\cite{petsc-efficient, petsc-user-ref, petsc-web-page} for the distributed implementation of iPI methods. PETSc is an open-source C library that, while originally designed to efficiently solve large-scale partial differential equations (PDEs), provides a rich set of general-purpose, optimized solvers for linear systems and a wide range of linear algebra data types and routines. These, along with PETSc's distributed memory parallelism capabilities, allow madupite to scale computations beyond a single CPU, making it possible to solve large-scale MDPs in a fully distributed manner.
\subsubsection{Distributed Memory Parallelism}\label{subsubsec: distributed_mem_par}
Madupite inherits from PETSc a distributed memory parallelism. This differs fundamentally from the fork-join thread model, where threads share the same memory space. Instead, distributed parallelism involves executing multiple instances of a program, each with its own separate memory space. Communication between these instances is handled using the Message Passing Interface (MPI) standard~\cite{mpi_paper}. Each instance of the program is called a rank, identified uniquely by an integer $\rho \in \left\{0,\dots, R-1 \right\}$, where $R$ is the total number of processes in the computation.
The ranks operate within an MPI communicator, and since ranks abstract away from the specifics of the CPU architecture, the same madupite code can run in a fully distributed manner across multiple cores, sockets, or even different machines. This makes our solver highly adaptable to a range of hardware configurations without requiring the minimal change on the code side from the user. However, to minimize communication costs, it is essential to organize ranks based on a physical-proximity logic: ranks that need to communicate frequently should be physically close to one another in order to reduce communication latency and therefore enhance the overall solver performance.

\subsubsection{Data Structures and Layouts}\label{subsubsec: datastructures_layouts}
Madupite uses \textit{Mat}, \textit{Vec} and \textit{IS} data structures of PETSc to store quantities of the MDP as well as for implementing iPI methods. In particular, \textit{Mat} denotes a generic matrix type and in madupite is deployed in compressed sparse row (CSR) format for storing the transition probabilities, and in dense format for storing the stage-cost values. Even if more specialized data structures, such as sparse block matrices or symmetric matrices, are also available in PETSc, we opt for the more generic, yet less optimized, {MatMPIAIJ} data structure as madupite should be able to handle any MDP, independently from its underlying structure. The latter is a distributed sparse matrix format optimized for parallel computation involving matrices with a sparse structure. Obviously, when possible, tailoring more the data structure to the geometry of the underlying MDP could result in better performance, but for the time being we leave this extension for future work. Since there is no support for data structures with more than 2 dimensions in PETSc, we had to select a proper 2-dimensional layout and flatten the 3-dimensional transition probability tensor $P\in \mathbb{R}^{n\times n\times m}$. Different layouts were taken into considerations: for instance flattening to a $n\times nm$- matrix by horizontally stacking the transition probabilities of every action index. This layout would have been the natural choice considering the fact that matrices are distributed by groups of rows in PETSc and that, for communication efficiency, all data $P(s,\cdot,\cdot)$ associated to the same state $s$ should ideally be stored in the same rank. This allows, for instance, local computation of portions of the greedy policy without need for extra communication among ranks. By extensive profiling and testing, it turned out that row access is much more efficient than column access due to the underlying CSR format for the sparse matrices. This strongly advocates against the choice of this layout given the repeated operations of greedy policy extraction (step 3 of Algorithm~\ref{alg:iPI}) and computation of its associated transition probability matrix (step 4 of Algorithm~\ref{alg:iPI}). We therefore resorted to a row-stacked layout, which results into the following $nm\times n$ flattened tensor 
\begin{equation}\label{eq:flattened_tensor}
    \begin{bmatrix}
        P(1,1,1) & P(1,2,1) & \dots & P(1,n,1)\\
        P(1,1,2) & P(1,2,2) & \dots & P(1,n,2)\\
        \vdots & \vdots & \ddots & \vdots \\
        P(1,1,m) & P(1,2,m) & \dots & P(1,n,m)\\
        P(2,1,1) & P(2,2,1) & \dots & P(2,n,1)\\
        \vdots & \vdots & \ddots & \vdots \\
        \vdots & \vdots & \ddots & \vdots \\
        P(n,1,m) & P(n,2,m) & \dots & P(n,n,m)\\
    \end{bmatrix}\,.
\end{equation}
 Matrix~\eqref{eq:flattened_tensor} is distributed such that all $P(s,\cdot,\cdot)$ associated to the same state $s$ are stored in the same rank. Therefore in each rank we store a coherent chunk of consecutive rows from~\eqref{eq:flattened_tensor}. To compute the precise number of rows, we first compute the local number of states stored on each rank by keeping a balanced load over the ranks. Assuming that we distribute the $n$ states across $R$ ranks, then the local number of states for rank $\rho$ is computed as follow
\begin{equation}
    n_{\rho} = 
    \begin{cases}
       \lfloor n/R \rfloor +1 & \text{if } \rho < n \% R \\
       \lfloor n/R \rfloor & \text{else}
    \end{cases}
\end{equation}
where $\rho \in \left\{ 0,\dots, R-1 \right\}$. The local number of rows of~\eqref{eq:flattened_tensor} that is stored in rank $\rho$ is then $n_{\rho}m$.
For the stage-cost matrix we instead do not need to alter the original 2-dimensional layout and each rank stores its corresponding $n_{\rho}$ consecutive rows from this matrix.
\textit{Vec} denotes a vector of real scalars and it is used to store the iterates and interim results of iPI methods. Finally, in madupite we store policies as \textit{IS} vectors. This data structure is a vector of indexes and allows one to efficiently access specific elements in a matrix.

Finally, the parameters and data that define an MDP are stored in madupite as attributes of a general \textit{MDP} class.

\subsubsection{Implementation Details on Inexact Policy Iteration Methods}\label{subsubsec: implementation_details_iPI}
 From an implementation perspective, an inexact policy iteration method can be broken down into three fundamental steps: extracting the greedy policy (step 3 in Algorithm~\ref{alg:iPI}), evaluating the coefficient matrix and forming the preconditioner (step 4 and 5 in Algorithm~\ref{alg:iPI}), and approximately solving the policy evaluation linear system (steps 6-12 in Algorithm~\ref{alg:iPI}). Below, we briefly discuss the key implementation choices for each step.

Extracting the greedy policy involves solving an arg-minimization problem over the action space. There are several possible approaches, each differing in terms of memory allocations and overall memory usage. We explored two main strategies for step 3 in Algorithm~\ref{alg:iPI}.

The first approach involves looping over the action space. During the \(a\)-th iteration, \(P_{\pi}\), \(g_{\pi}\), and \(g_{\pi} + \gamma P_{\pi}\) are computed with \(\pi = [a, \dots, a]\). The action \(a\) is stored if the resulting value is smaller than previously computed values. This method is memory-efficient because it only tracks the minimum value, resulting in a memory complexity of \(\mathcal{O}(n)\). However, since the CSR format does not allow changing the structure of a matrix after allocation, this approach requires reallocating \(P_{\pi}\) for a total of \(m\) times, leading to poor performance in testing.

The second approach we evaluated involves directly computing the product \(PV\), which yields a vector of size \(nm\). This vector is reshaped into a dense matrix of size \(n \times m\), scaled by \(\gamma\), and added to \(g\). A greedy policy is then obtained by performing the arg-minimization on this matrix. Although this method has a higher memory complexity of \(\mathcal{O}(mn)\)—with the factor \(m\) being significant in settings with large action spaces—it requires only a single matrix allocation. Since the number of memory allocations was identified as the primary performance bottleneck, we opted for this second approach, which proved to be more efficient.

The inexact policy evaluation step in iPI entails approximately solving the linear system in \eqref{eq:bellman_eq_pi}, where \(\pi\) is the current greedy policy computed at step 3. To achieve this, we utilize PETSc's KSP class, which offers a variety of Krylov subspace iterative methods, including the generalized minimal residual method (GMRES), the biconjugate gradient stabilized method (BiCGStab), the transpose-free quasi-minimal residual method (TFQMR), Richardson’s method, and the conjugate gradient method (CG)~\cite{saad}. Madupite allows users to employ most of these methods, provided they are compatible with PETSc's MATMPIAIJ matrix class, which, as described in 2), is the distributed sparse matrix format optimized for parallel computation used by madupite. 

Consider the general inexact policy iteration method described in Algorithm~\ref{alg:iPI} where step 8 is performed using a Krylov subspace method. In this case, the algorithm does not require direct access to the elements of the coefficient matrix \(I - \gamma P_{\pi}\). Therefore, instead of forming this matrix explicitly from \(P_{\pi}\), we opt for a \textit{matrix-free} approach by using a \textit{matrix shell}. This allows for a logical representation of the matrix without allocating or constructing it. However, when the SOR preconditioner is selected in step 5, explicit access to matrix elements is necessary, which requires madupite to explicitly allocate and compute the coefficient matrix, which typically leads to reduced efficiency. 

Regarding the stopping condition for the inner loop in step 8 of Algorithm \ref{alg:iPI}, while the right-hand side is readily available from the outer loop computations, the left-hand side must be recomputed at each inner iteration. This adds considerable overhead, especially when a large number of inner iterations is needed. To optimize performance, we replace the original stopping condition with a more stringent criterion that, instead of the infinity-norm, leverages the 2-norm of the linear residual. The latter is provided already by the solver at each step as it is deployed for its internal computations, allowing for a significative reduction of the computational overhead and ultimately enhancing the solver's efficiency.



\subsection{Python API \& Deployment}\label{subsec: pythonAPI_deployment}

Madupite is a fully-featured C++20
library and, by leveraging the nanobind binding library~\cite{nanobind}, offers an equivalent API
in Python that can be installed as a package using pip.
Madupite's Python API allows users to simply deploy madupite from Python, without the need to write complex C++ code, but also without loosing in functionalities, performance and flexibility in terms of customization. We will now briefly discuss the main principles of madupite's Python API. For more details as well as tutorials and examples, we refer the users to madupite's documentation~\cite{madupite_doc}.

Madupite can be imported and used as any other Python package. Though, if you want to run your code in a parallel/distributed set up, the code should be contained inside a Python main function, as this guarantees a correct finalization of the MPI ranks.

MDPs are implemented in madupite via a Python class that we name MDP.
Creating a specific MDP in madupite therefore boils down to creating an instance of madupite's MDP class and setting its attributes to the madupite's objects containing the information for the specific MDP that we aim at solving. This can be done via the \codeword{setStageCostMatrix} and \codeword{setTransitionProbabilityTensor} methods of the MDP class, which take as input the madupite's objects containing the stage-cost matrix and the transition probability tensor information, respectively. All other features, such as the discount factor value, the optimization mode and the hyperparameter configuration for the iPI method, can be set as attributes of the MDP instance via its method \codeword{SetOption}. We report in Table~\ref{table1} a list of some of the main available options and for each of those we report type, default value and we indicate to which hyperparameter in Algorithm~\ref{alg:iPI} they correspond to. For each choice of the inner solver, there are also solver-specific options supported by PETSc and those can be set in madupite with the same syntax. 
A complete list of supported inner solvers with their specific options is available at \url{https://petsc.org/release/manualpages/KSP/KSPType/}. For instance, it is possible to customize Richardson's method by selecting the value of the step-size via the \codeword{-ksp_richardson_scale} option as described at \url{https://petsc.org/release/manualpages/KSP/KSPRICHARDSON/}. Even if madupite employs inexact policy iteration methods, these approaches are sufficiently general to encompass the core dynamic programming algorithms as well. It is indeed possible to recover value iteration, policy iteration and optimistic policy iteration by appropriately selecting the hyperparameter configuration for the solver via the options. We illustrate that in Table~\ref{table2}, where we describe how to set the hyperparameters of the solver via the options in order to obtain the classical dynamic programming methods. In particular, we show how to recover Policy Iteration (PI); Value Iteration (VI); $\alpha$-Value Iteration~\cite{gargiani_2022, gargiani_extended_2022}, which we re-name $\beta$-Value Iteration ($\beta$-VI) to avoid notation overlaps with the stopping condition parameter; Optimistic Policy Iteration (OPI)~\cite{DB_book}, where we use OPI$(W)$ to denote OPI when $W$-inner iterations are used in the inner loop; Gauss Seidel-Value Iteration (GS-VI) and Jacobi-Value Iteration (J-VI)~\cite{gargiani_minibatch}.  
In order to set the MDP's data as attributes of the instance of the MDP class, there is need for the user to create appropriate madupite's objects before. This can be done via apposite madupite's functions.
In particular, madupite supports both reading data from compliant binary files with the \codeword{Matrix.fromFile} madupite function; or generating them in real time via plain Python functions with the \codeword{createStageCostMatrix} and \codeword{createTransitionProbabilityTensor} madupite functions. The latter approach requires the user to write two plain Python functions, one for the stage-cost and one for the transition probabilities, which are then passed as input arguments to the respective madupite function. Each rank will then simulate the functions for their local portion of states, generating and storing the data in a fully distributed manner. In this setting it is strongly advisable to deploy pre-allocation as it may greatly enhance performance. 
It is also possible to combine the two approaches. For instance, assume the stage-cost is designed via a specific function from the user while transition data are coming from empirical simulations which have been stored into a file. In this scenario, the second approach can be used to generate a madupite's object for the stage-cost, while the first approach is used to generate a madupite's object for the transition probability matrix.
Providing a full tutorial is beyond the scope of this paper; our aim here is to offer a general overview of the solver and its deployment.
We therefore refer the reader to~\cite{madupite_doc} for a detailed implementation guidance. 
\begin{table*}
\begin{center}
\begin{tabular}{||c c c c||} 
 \hline
 \textbf{name} & \textbf{type} & \textbf{default} & \textbf{parameter} \\ [0.5ex] 
 \hline\hline
 \codeword{-max_iter_pi} & INT & 1000 & $N_o$ \\ 
 \hline
 \codeword{-max_iter_ksp} & INT & 1000 & $N_i$ \\
 \hline
 \codeword{-atol_pi} & DOUBLE & 1e-8 & $tol$ \\
 \hline
 \codeword{-alpha} & DOUBLE & 1e-4 & $\alpha$ \\
 \hline
 \codeword{-ksp_type} & STRING$^*$ & gmres & IterativeSolver \\
 \hline
 \codeword{-pc_type} & STRING$^{\dagger}$ & none & $D_{k}$ \\  
 \hline
\end{tabular}
\end{center}
\caption{In this Table we report the values of the main options available in madupite. STRING$^*$ and STRING$^{\dagger}$ are used to compactly denote that the value should be a string and the feasible values can be found at \url{https://petsc.org/release/manualpages/KSP/KSPType/} and \url{https://petsc.org/release/manualpages/PC/PCType/}, respectively.}
\label{table1}
\end{table*}

\begin{table*}
\begin{center}
\begin{tabular}{||c c c c||} 
 \hline
 \textbf{DP-algorithm} & {\codeword{ksp_type}} &  \codeword{pc_type} & extra inner solver options \\ [0.5ex] 
 \hline\hline
 VI & richardson &  none & \codeword{-ksp_max_it 1}\\ 
 \hline
 PI & richardson & svd &  \\
 \hline
 OPI(W) & richardson & none & \codeword{-ksp_max_it W}\\
 \hline
 $\beta$-VI & richardson & none & \codeword{-ksp_max_it 1 -ksp_richardson_scale }$\beta$\\
 \hline
 GS-VI & richardson & sor & \codeword{-pc_sor_forward -ksp_max_it 1} \\
 \hline
 J-VI & richardson & jacobi & \codeword{-ksp_max_it 1}\\
 \hline
\end{tabular}
\end{center}
\caption{In this Table we illustrate how to set madupite's options to recover the main algorithms from the DP literature.}
\label{table2}
\end{table*}

\section{Performance Evaluation}\label{sec: performance_evaluation}
This section presents a comprehensive evaluation of the performance of madupite solver. We begin in Subsection~\ref{sec: performance_evaluation}-\ref{subsec: performance_evaluation}.\ref{subsubsec:amdahlslaw} by analyzing its parallelizability and speedup capabilities, examining how well the solver can leverage parallel processing for enhanced efficiency. Next, in Subsection~\ref{sec: performance_evaluation}-\ref{subsec: performance_evaluation}.\ref{subsubsec:solvercustomization} we highlight the flexibility of madupite in customizing iPI methods, demonstrating how this adaptability contributes to faster convergence rates. In Subsection~\ref{sec: performance_evaluation}-\ref{subsec: performance_evaluation}.\ref{subsubsec:memorylimitations} we assess the scalability of madupite, evaluating its performance as the problem size grows. In Subsection~\ref{sec: performance_evaluation}-\ref{subsec: performance_evaluation}.\ref{subsubsec:comparisonSOTA} we benchmark madupite against SOTA solvers for MDPs. For this first set of benchmarks, we use artificial MDPs with randomly generated stage costs and transition probabilities for general performance evaluation. Following this, we showcase madupite’s practical effectiveness through two case studies from distinct application areas: epidemiology and classical control theory. Specifically, in Subsection~\ref{sec: performance_evaluation}-\ref{subsec:infectiondisease} we examine the control of viral disease spread, which results in a large-scale MDP with discrete state and action spaces; in Subsection~\ref{sec: performance_evaluation}-\ref{subsec:pendulum} the optimal control of a classic inverted pendulum system, which we model as a continuous-state MDP that requires discretization for solution, and in Subsection~\ref{sec: performance_evaluation}-\ref{subsec: maze} we demonstrate the use of madupite to solve navigation tasks in two-dimensional mazes of increasing size.
\subsection{General Performance Evaluation}\label{subsec: performance_evaluation}

\subsubsection{Amdahl's Law}\label{subsubsec:amdahlslaw}
Let $p\in [0,1]$ denote the parallelizable portion of our code. According to Amdahl's law~\cite{amdahl1967}, the theoretical speedup in execution for a fixed workload when $R$ ranks are deployed instead of a single one is given by
\begin{equation}\label{eq:amdahlslaw}
    S(R) = \frac{1}{1-p + \frac{p}{R}}\,,
\end{equation}
where $S(R)$ denotes the speedup achieved with $R$ ranks.

\begin{figure}
    \centering
    \includegraphics[width=\linewidth]{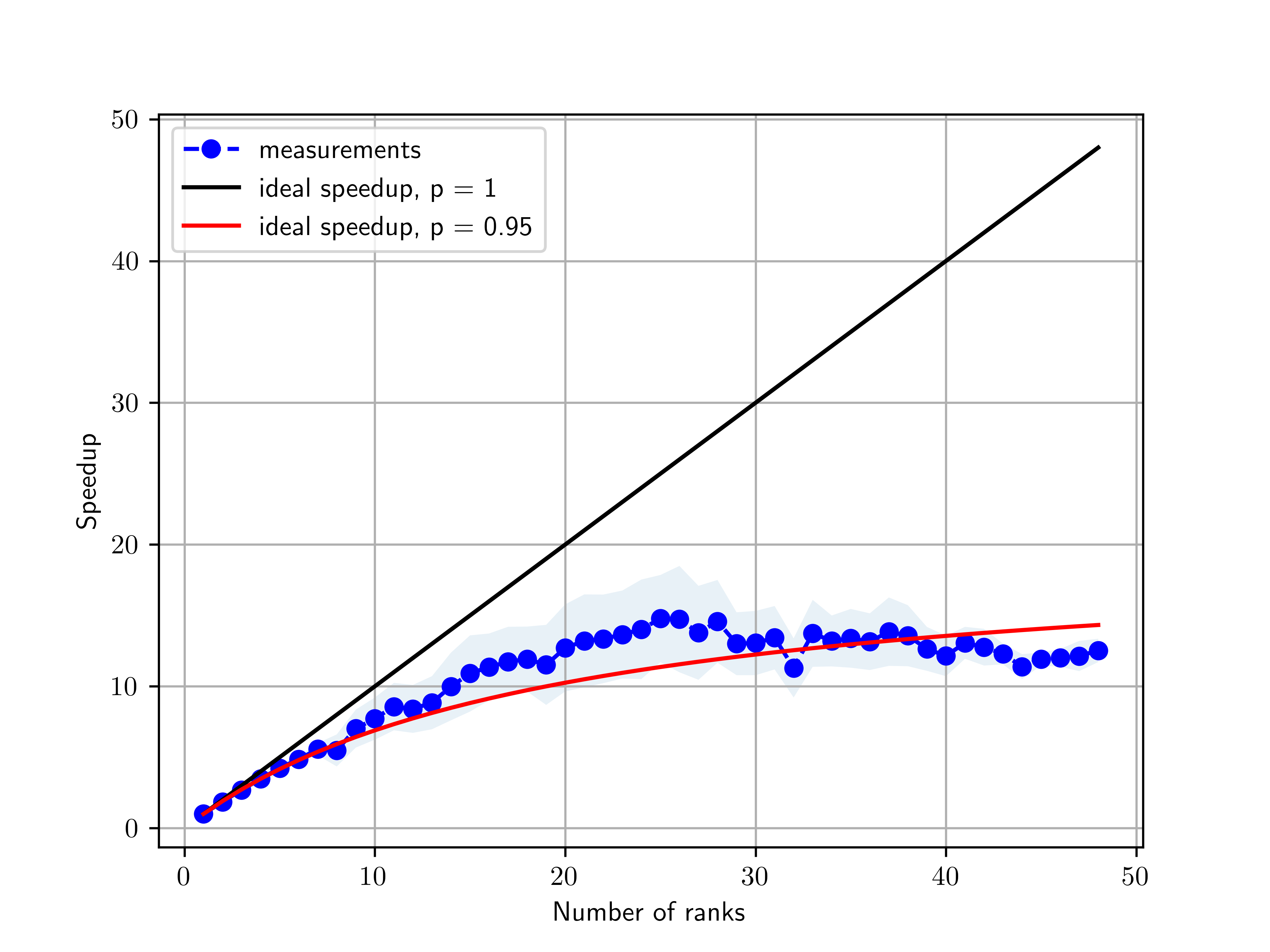}
    \caption{Measured speedup versus number of ranks. The blue dotted line with markers shows the median speedup observed across multiple runs, with the shaded blue region indicating the standard deviation. The black line represents the ideal linear speedup assuming a fully parallelizable code (p = 1), while the red line corresponds to the ideal speedup under Amdahl’s Law for a 95\% parallelizable workload (p = 0.95).}
    \label{fig:amdalhslaw}
\end{figure}

The goal of this benchmark is to deploy Equation~\eqref{eq:amdahlslaw} and runtime measurements to estimate the portion of parallelizable code in madupite as well as the maximum achievable speedup. For that we deploy an artificial MDP with 100.000 states, 100 actions and a discount
factor of 0.69. In particular, we sample the stage costs and transition probabilities uniformly. For the latter, we also limit the reachable set of states for each state-action pair to induce sparsity for a total of 1.5 billion non-zero entries. We use GMRES as inner solver, $\alpha = 0.0001$ and we run all benchmakrs on Euler HPC cluster~\cite{euler_cluster}. For each value $\rho \in \left\{1,\dots, 48 \right\}$ we run 10 instances and compute the median runtime. The measured runtimes are also used with~\eqref{eq:amdahlslaw} to estimate the parameter $p$. As an estimator, we use the minimizer of the absolute error, which offers greater robustness to outliers compared to the least-squares method. Our experiments returned a value of 0.95, indicating that 95$\%$ of our code is parallelizable.
The remaining 5$\%$ of our code is inherently non-parallelizable, primarily due to synchronization and communication between ranks. Substituting this value back into Equation~\eqref{eq:amdahlslaw} and taking the limit as $R\rightarrow\infty$ reveals that the maximum achievable speedup is approximately 20. In Figure~\ref{fig:amdalhslaw}, we plot with a dotted blue line the median runtime as a function of the number of ranks, with the standard deviation indicated by a shaded blue area. The red and black lines represent the ideal speedup curves for code that is 95\% and 100\% parallelizable, respectively. 

\subsubsection{Solver Customization}\label{subsubsec:solvercustomization}
As established in the literature, the discount factor significantly affects the convergence rate of value iteration: higher discount factors lead to slower convergence, increasing the number of iterations needed to reach a solution with acceptable accuracy (see among others~\cite{DB_book, gargiani_2023, gargiani_iPI}). Additionally, as discussed in~\cite{gargiani_iPI}, the performance of iPI methods is strongly influenced by the convergence characteristics of the inner solver. In high discount factor settings, value iteration suffers from such an exacerbatingly slow convergence rate that completely offsets the benefit of its low per-iteration computational cost. As a result, solution methods based on value iteration, like OPI, exhibit poor runtime performance.

\begin{figure}
    \centering
    \includegraphics[width=\linewidth]{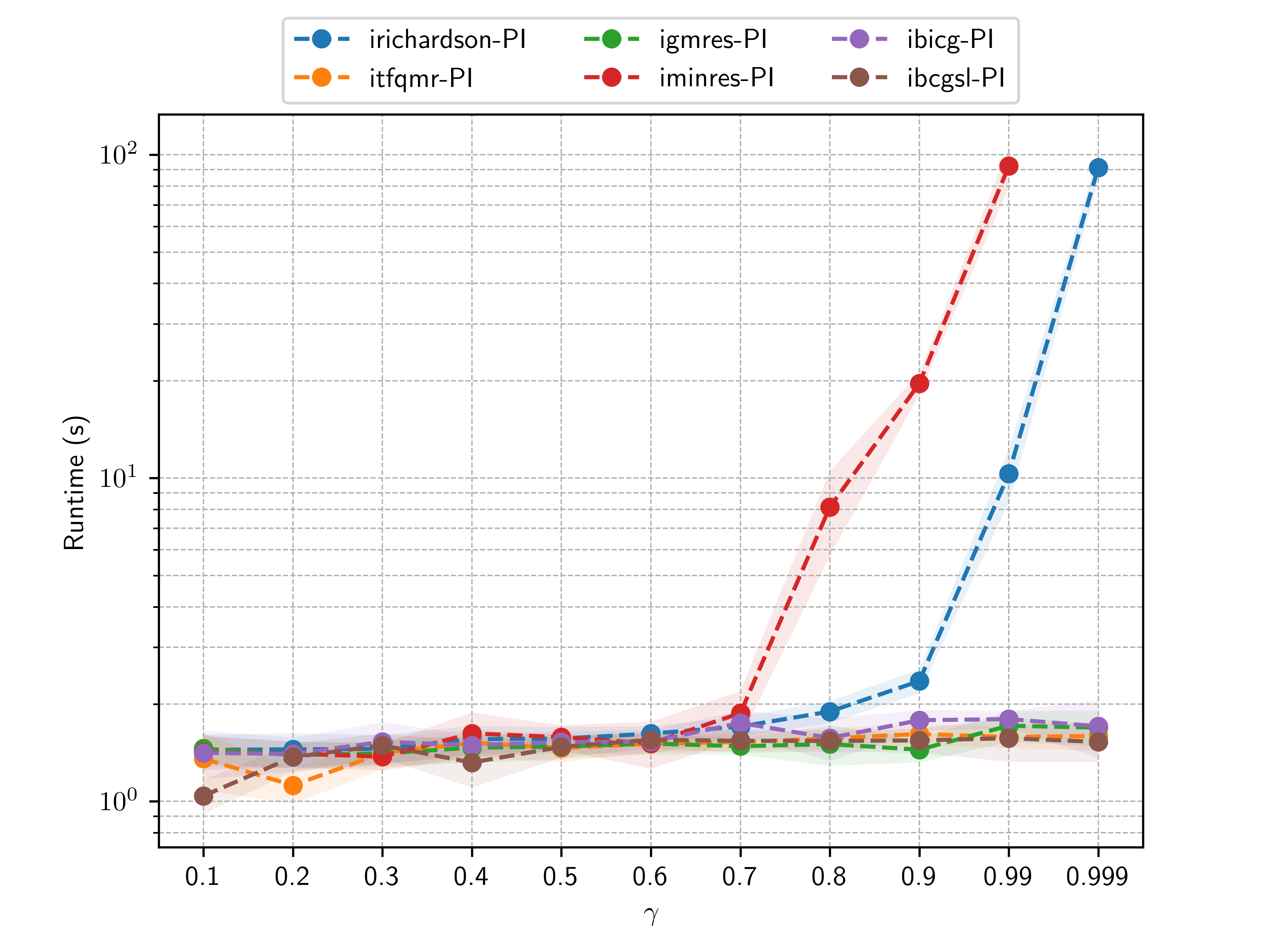}
    \caption{Median runtimes in seconds over 10 runs versus discount factor. Shaded areas represent standard deviation and different colours are associated to iPI methods with different inner solvers.}
    \label{fig:innersolvers}
\end{figure}

To the best of our knowledge, madupite is the only existing MDP solver that enables users to customize the solution method, allowing them to select from a wide range of inner solvers, including sophisticated ones such as GMRES and BiCGSTAB. These alternatives have demonstrated superior performance in high discount factor settings compared to traditional value iteration~\cite{gargiani_iPI}. The objective of these benchmarks is to highlight the substantial benefits of selecting an appropriate inner solver, underscoring the importance of having the flexibility to customize this choice — a feature unique to madupite. As a scalable variant of PI for large-scale settings, other existing MDP solvers typically rely only on OPI, which is equivalent to iPI using Richardson’s method as the inner solver. However, this reliance on OPI poses a significant performance limitation, particularly in scenarios with high discount factors. This is the core insight we aim to convey through the following set of numerical evaluations. For this set of experiments, we use the same artificial MDP setup as in~\ref{sec: performance_evaluation}-\ref{subsec: performance_evaluation}.\ref{subsubsec:amdahlslaw}. We increase gradually the discount factor and study its effect on the convergence of iPI methods when different inner solvers are selected. In order to evaluate purely the impact of the inner solver, we run the benchmarks in single-core. In Figure~\ref{fig:innersolvers}, we present the median runtime over 10 independent runs, shown using dashed lines, with shaded areas representing the standard deviation and we use i$\langle innersolver \rangle$-PI to indicate iPI with $\langle innersolver \rangle$ as inner solver. As illustrated in Figure~\ref{fig:innersolvers}, both irichardson-PI and iminres-PI exhibit poor performance at high discount factors. In contrast, the other methods remain largely unaffected by the increase in the discount factor and demonstrate significantly better performance. 

\subsubsection{Beyond The Memory Limitations of a Single-Laptop}\label{subsubsec:memorylimitations}
A key innovation of madupite is its ability to scale the storage and solution of MDPs to sizes that exceed the capacity of a single computer. Typically, a modern laptop has only 4 to 6 GB of available physical memory, which constrains the size of state and action spaces that can be stored and solved exactly. By distributing both storage and computation, madupite establishes a new state-of-the-art, enabling the use of high-performance computing clusters to solve exactly larger MDP instances. 
\begin{figure}
    \centering
    \includegraphics[width=\linewidth]{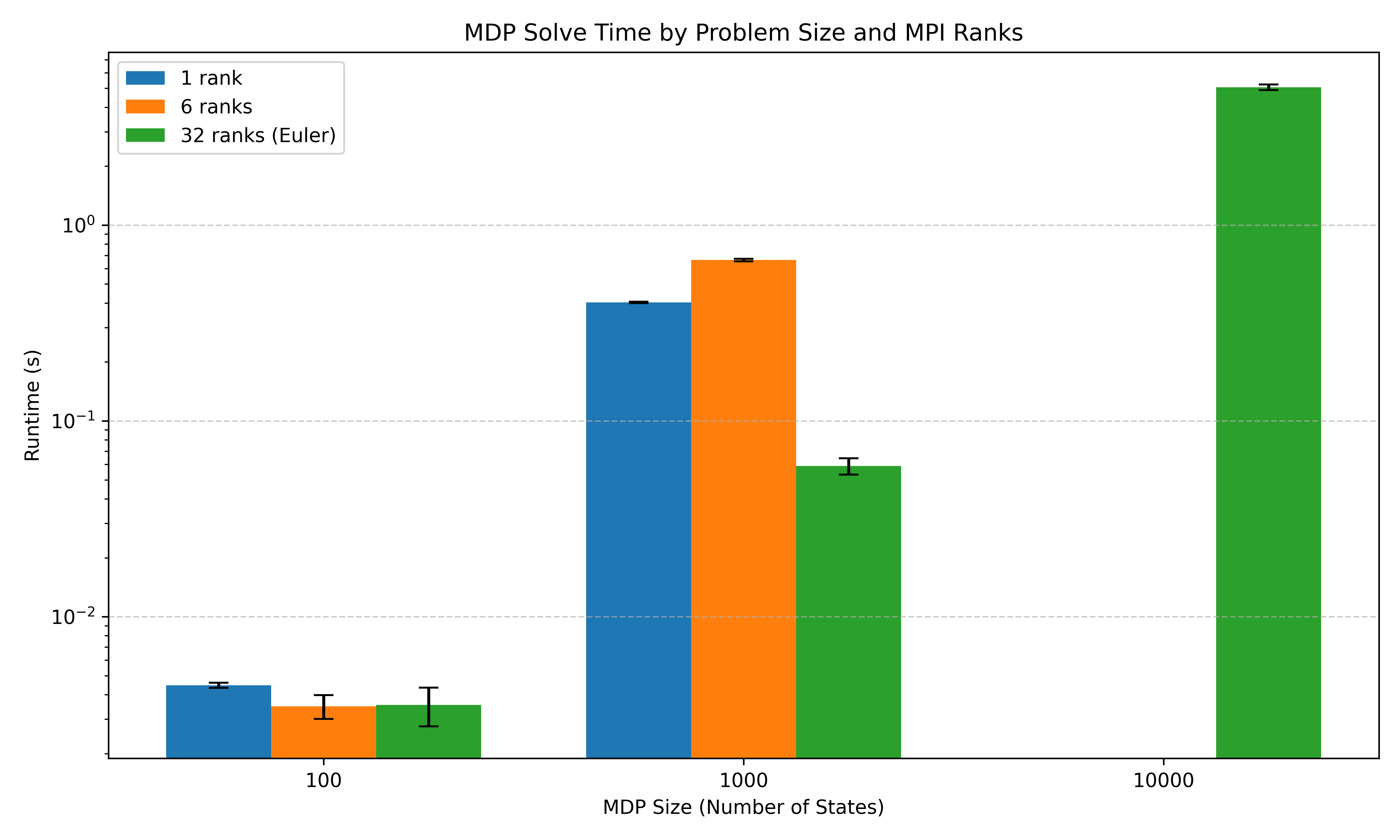}
    \caption{Mean runtime (in seconds) over 10 runs versus state space size. Blue and orange bars represent runs using 1 and 6 cores, respectively, on an Intel Core™ i7-10750H laptop CPU, while green bars correspond to runs using 32 cores on the Euler cluster.}
    \label{fig:memory_runtime}
\end{figure}
For these benchmarks, we use artificially generated dense MDPs with a fixed action space of size 100 and a discount factor of 0.99. We progressively increase the number of states to evaluate how well madupite scales in both memory and computation. The results, shown in Figure~\ref{fig:memory_runtime}, report the mean and standard deviation of the solver’s runtime.

Blue and orange bars represent runs on a single core and 6 cores, respectively, on an Intel Core™ i7-10750H laptop CPU. The green bars correspond to runs with 32 cores on the Euler HPC cluster~\cite{euler_cluster}. As the size of the MDP increases, we observe that the runtime on the laptop grows significantly and eventually fails due to memory limitations. When $n = 1000$, the transition probability matrix alone demands around 74GiB of memory, significantly exceeding the RAM available on typical laptops and underscoring their unsuitability for handling such large-scale MDPs. In contrast, madupite can solve much larger instances on the cluster by distributing both storage and computation across multiple nodes. 

An interesting trend emerges for smaller problem sizes: on the laptop, increasing the number of ranks from 1 to 6 yields little to no performance gain, and in some cases, even degrades performance. This is primarily due to the overhead of parallel execution and contention for shared hardware resources, which outweigh the computational benefits for small workloads. In contrast, for larger MDPs, the parallel solver scales much more effectively. The 32-rank runs on the Euler HPC cluster~\cite{euler_cluster} achieve significantly faster runtimes and are capable of solving problem instances that exceed the memory capacity of a single laptop.

This benchmark highlights two key points: (1) naive parallelism does not guarantee speedup for small problems, and (2) madupite enables solving large MDPs that would otherwise be infeasible on a single machine, by leveraging distributed memory and computation.

\subsubsection{Comparison with SOTA MDPs Solvers}\label{subsubsec:comparisonSOTA}

Although madupite is primarily designed for solving large-scale MDPs on high-performance computing clusters — leveraging numerous ranks to distribute storage and computation —  in this section we empirically demonstrate its competitive performance even when deployed in single-core or multi-cores shared-memory. In particular, we compare madupite against pymdptoolbox~\cite{pymdptoolbox} and mdpsolver~\cite{MDPsolver}, both recognized as state-of-the-art solvers for MDPs.
For that, we use pymdptoolbox, mdpsolver, and madupite to solve a randomly generated MDP with 1000 states, 500 actions, and a discount factor of 0.999. The tolerance is uniformly set to $10^{-6}$ for all solvers. While both pymdptoolbox and mdpsolver rely on OPI as solution method, we deploy madupite with GMRES as inner solver and $\alpha = 0.001$. The benchmarks are conducted on an Intel(R) Core(TM) i7-10750H CPU @ 2.60GHz, with mdpsolver running in parallel mode. 

Table~\ref{table_SOTA} presents the average runtime across 10 independent runs. In the single-core setting, madupite outperformed its competitors, achieving speedups of $\times 2.05$ and $\times 1.95$ compared to pymdptoolbox and mdpsolver, respectively. These speedups increased to $\times 2.88$ and $\times 2.7$ when madupite's multi-core capabilities were utilized, running with 4 ranks instead of 1.
 
\begin{table*}
\begin{center}
\begin{tabular}{||p{3.3cm} p{2.3cm} p{2.3cm} p{2.4cm}||}
 \hline
 \multicolumn{4}{||c||}{\textbf{Runtime [s]}} \\
 \hline
  \,\,\,\texttt{pymdptoolbox} & \,\,\texttt{mdpsolver} & madupite(1) & madupite(4) \\
 \hline
  2.542  & 2.384 & 1.211 & 0.881 \\
 \hline
\end{tabular}
\end{center}
\caption{Average runtime over 10 independent runs. We use madupite$(R)$ to denote the deployment of $R$-ranks. The \texttt{mdpsolver} is used in parallel mode.}
\label{table_SOTA}
\end{table*}

\subsection{Infectious Disease Models}\label{subsec:infectiondisease}
\begin{figure*}
    \centering
    \includegraphics[width=0.8\linewidth]{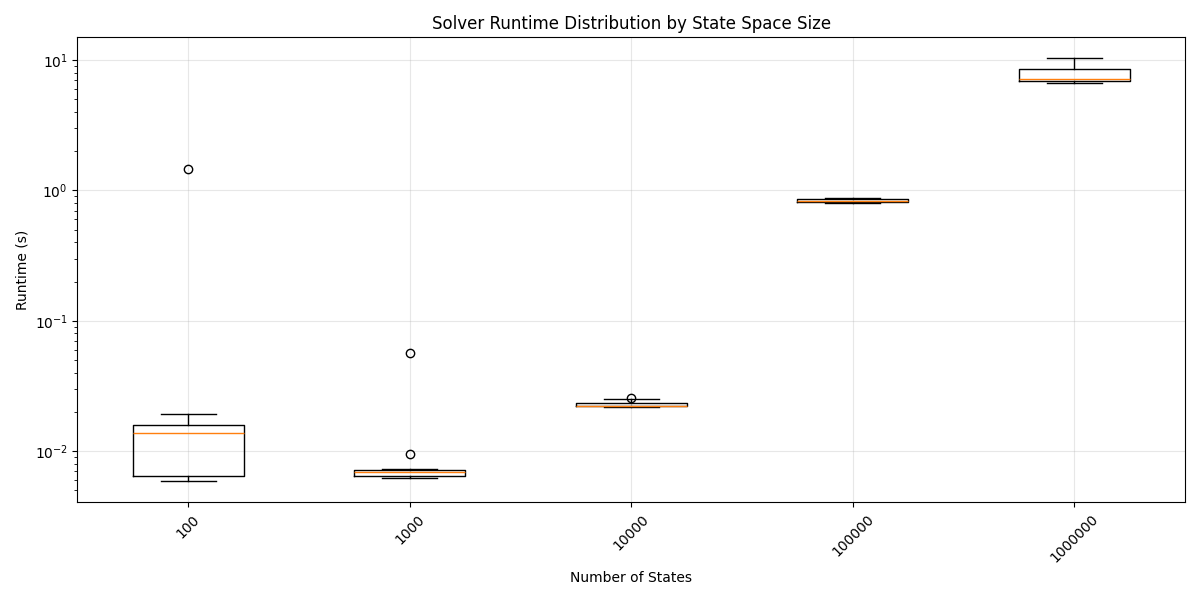}
    \caption{Runtime versus number of states when madupite is used to solve MDPs arising from dynamic SIS models. The benchmarks are run on 32 cores of the Euler cluster.}
    \label{fig:sis_bench}
\end{figure*}

We consider a Susceptible-Infectious-Susceptible (SIS) model to describe the evolution of an infectious disease with no immunity conferred by the previous infection~\cite{hethcote}. As suggested by its name, this model is characterized by two classes: the susceptible and the infectious class. Individuals move from the susceptible class to the infectious class and then back to the susceptible class upon recovery.
As in~\cite{sis_model}, we assume that the infectious period is fixed and equal to 1; that is, an individual that is infectious at time $t$ remains infectious over the interval $[t, t+1]$, but she/he will recover upon diagnosis and effective treatment at time $t+1$, returning to the susceptible class. We denote with $s(t)$ and $i(t)$ the number of individuals in the susceptible and infectious class at time $t$, respectively.   Since we consider populations with a fixed-size, then
\begin{equation}\label{eq: SIS_dynamics}
s(t) + i(t) = N \quad \forall \, t\geq 0\,.
\end{equation}
From~\eqref{eq: SIS_dynamics} we obtain that only the information of a single class are needed to describe the state of the disease and we choose $s(t)$ to be the state of the model. From now on, we drop the dependency on time, unless needed.
We extend the static model in~\cite{sis_model} by adding a dynamic component which allows one to derive public health policies. In particular, we introduce actions, an action-state dependent stage-cost and action dependent distribution of the driving event. For the actions, we propose $\mathcal{A} = \mathcal{A}_1 \times \mathcal{A}_2$ as action set, where $\mathcal{A}_1 = \left\{ 0,1,2,3,4 \right\}$ are the levels of hygiene measures and $\mathcal{A}_2 = \left\{ 0, 1, 2, 3 \right\}$ are the levels of social distancing imposed by the public health authorities. Considering influenza as an example, levels of actions in increasing order in the first set could correspond to \textit{no measures}, \textit{frequent hand washing and disinfection}, \textit{mandatory surgical masks}, \textit{mandatory FFP2 masks} and \textit{mandatory full body protection}, respectively. Levels of actions in increasing order in the second set could correspond to \textit{no social restrictions}, \textit{mandatory social distancing}, \textit{restaurant and store closure} and \textit{full lockdown}, respectively. As stage-cost, we consider the following multi-objective cost function
\begin{equation*}
g(s, a) = w_f c_f(a) - w_q c_q(a) + w_h c_h(s) \,,
\end{equation*}
where $w_f,\,w_q,\,w_h \geq 0$ are weights, $c_f:\mathcal{A}\rightarrow \mathbb{R}$ captures the financial cost of the hygiene and social measures, $c_q :\mathcal{A}\rightarrow [0,1]$ assigns to each action a quality of life score, and $c_h:\mathcal{S}\rightarrow \mathbb{R}$ maps the number of infected people to the medical cost incurred for their treatments. In particular, for the benchmarks we use $w_f = 1$, $w_q = 0.1$ and $w_h = 2$, $c_f(a) = c_{f,\text{HM}}(a_1) + c_{f,\text{SD}}(a_2)$, $c_q(a) = c_{q,\text{HM}}(a_1)\cdot c_{q,\text{SD}}(a_2) $ and $c_h(s) = (N-s)^{1.1}$. 
The driving event from the susceptible class to the infectious class is the random variable $I(t)$, which represents the number of new infections occurring during the interval $[t, \,t+1]$. As in~\cite{sis_model}, we model the probability mass function for the driving event as a binomial distribution 
\begin{equation*}
P\left(I(t) = i \,\vert\, s\right)=
\begin{cases}
\binom{s}{i} q(s,a)^{i} (1-q(s,a))^{s-i} ,\quad 0\leq i \leq s  \\
\,\, 0,\quad \text{else},
\end{cases}
\end{equation*}
where $q : \mathcal{S}\times\mathcal{A} \rightarrow [0,1]$ with $q(s,a) = 1 - \exp\left({-\lambda(a)\beta(s)\psi(a)}\right)$ is the overall probability that a susceptible person becomes infected. The latter is a function of $\psi(a)$, $\beta(s)$ and $\lambda(a)$, which are, respectively, the probability that a susceptible person becomes infected upon contact with an infectious individual, the probability that the next interaction of a random susceptible person is with an infectious person, and the contact rate. These parameters are specific of the considered infectious disease. For the benchmarks, we use $\psi(a) = \psi(a_1)$ and $\lambda(a) = \lambda(a_2)$ and $\beta(s) = 1 - s/N$. 
Finally, we consider $s(t) = N$ as an absorbing state since no infected individuals remain in the population. We refer to~\cite[Table 2]{gargiani_iPI} for details on the values of the parameters used for the benchmarks.
We set $\gamma = 0.99$, $\alpha=0.01$, $tol=1e-7$ and use GMRES as inner solver. In our experiments, we evaluate the performance of madupite in solving problems for populations of increasing size, up to 1 million individuals. The benchmarks are run on 32 cores of the Euler cluster~\cite{euler_cluster}. As shown in Figure~\ref{fig:sis_bench}, madupite handles even the largest instance of this problem with remarkable efficiency, requiring less than 10 seconds to compute the solution. 

\subsection{Inverted Pendulum}\label{subsec:pendulum}
Finally, we consider the inverted pendulum system, a standard benchmark~\cite{inverted_pendulum} extensively used in the last 50 years by researchers in different fields, from classical control to robotics and reinforcement learning~\cite{inverted_pendulum_RL}. The system consists of a rigid rod pivoted at its base, with its center of mass above the pivot point. The system requires active control to keep the mass balanced in the upright position as this equilibrium is inherently unstable. We refer to~\cite{mit_lecture} for a more detailed description of a general inverted pendulum system. The state of the system is described by the pair $(\theta(\tau),\,\dot{\theta}(\tau))$, where $\theta(\tau)$ is the angular position of the pendulum with respect to the vertical position and $\dot{\theta}(\tau)$ is its angular velocity. Both are real valued functions in continuous time and we denote with $\tau$ the continuous time variable. We act on the system via the torque $F(\tau)$, which we apply on the pivot point. The goal is to actively maintain its position upright. The system dynamic for a pendulum whose length and mass are both equal to one is described by the differential equation
\begin{equation}\label{eq:inverted_pendulum_dynamic}
\ddot{\theta}(\tau) + g \sin(\theta(\tau)) = F(\tau) \,,
\end{equation}
where $g=9.80665\,\,meter/sec^2$ is the gravitational acceleration. We consider the cost function
\begin{equation}
    g(\theta(\tau), \dot{\theta}(\tau), F(\tau)) = 2[(\theta(\tau) - \pi)^2 +  \dot{\theta}(\tau)^2] + F(\tau)^2 \,,
\end{equation}
comprising a quadratic penalty on the deviation of the state from the target position and the zero velocity, together with a quadratic penalty on the input. 

\begin{figure*}
    \centering
    \includegraphics[width=0.8\textwidth]{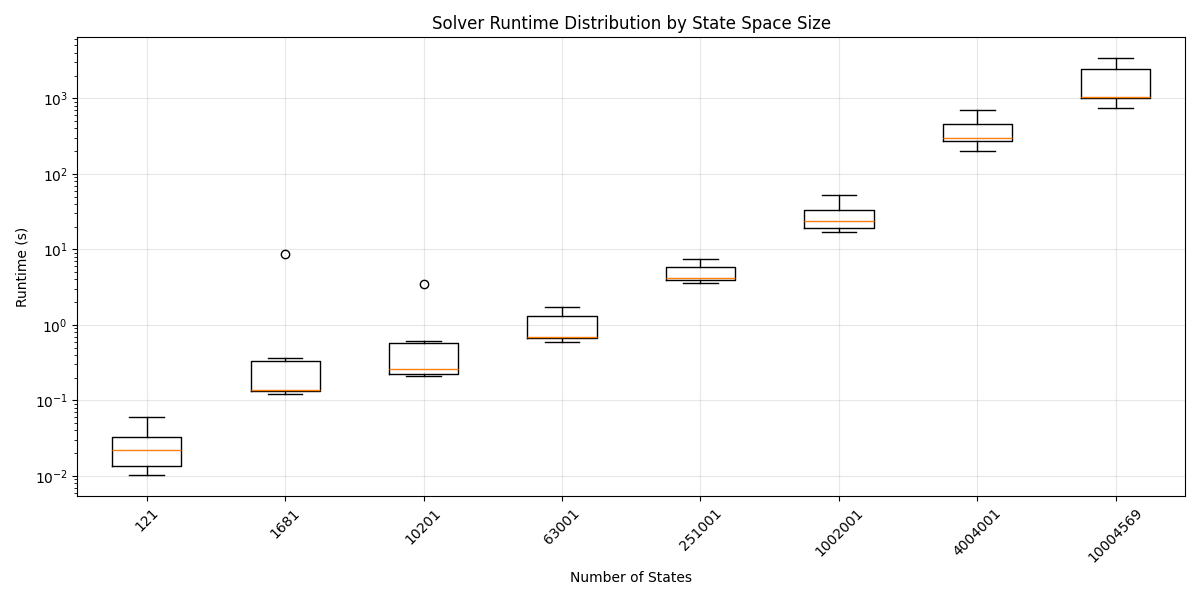}
    \caption{Runtime versus number of states when madupite is used to solve MDPs arising from the discretization of an inverted pendulum system. The benchmarks are run on 32 cores of the Euler cluster.}
    \label{fig:pendulum}
\end{figure*}

\begin{figure}
\centering
\begin{subfigure}[b]{0.45\textwidth}
   \includegraphics[width=1\linewidth]{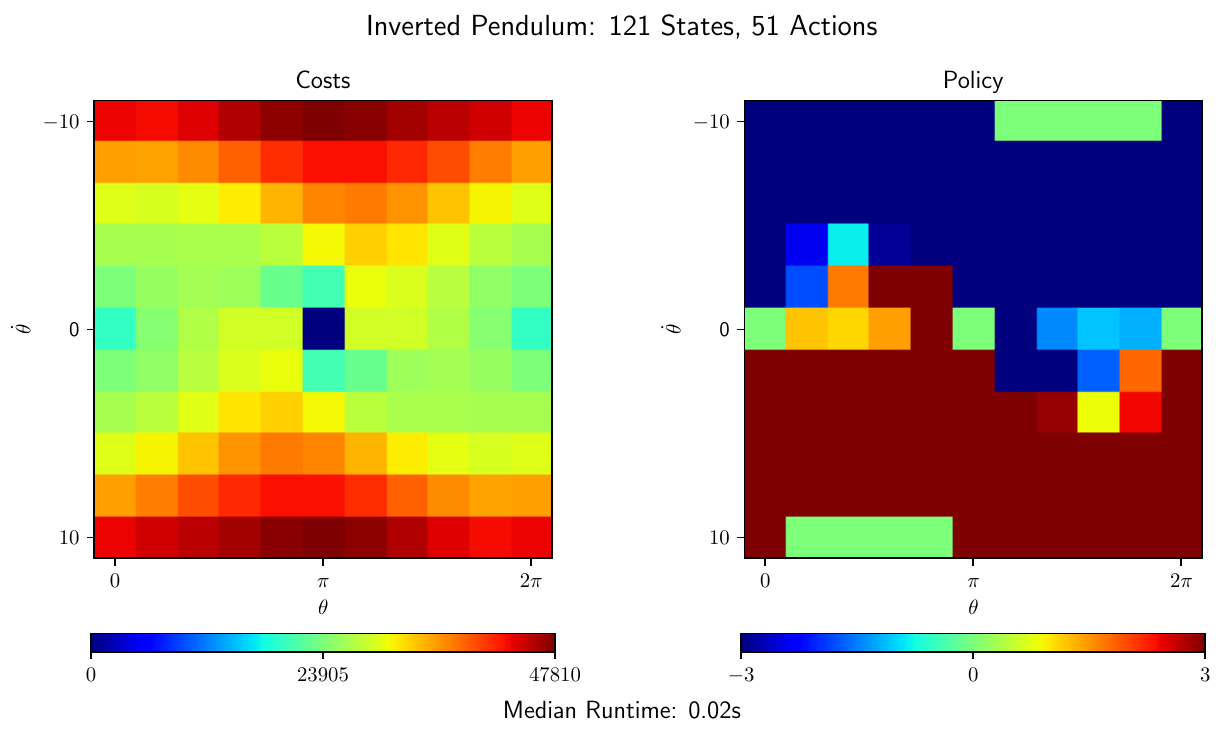}
   \caption{}
   \label{fig:low} 
\end{subfigure}

\begin{subfigure}[b]{0.45\textwidth}
   \includegraphics[width=1\linewidth]{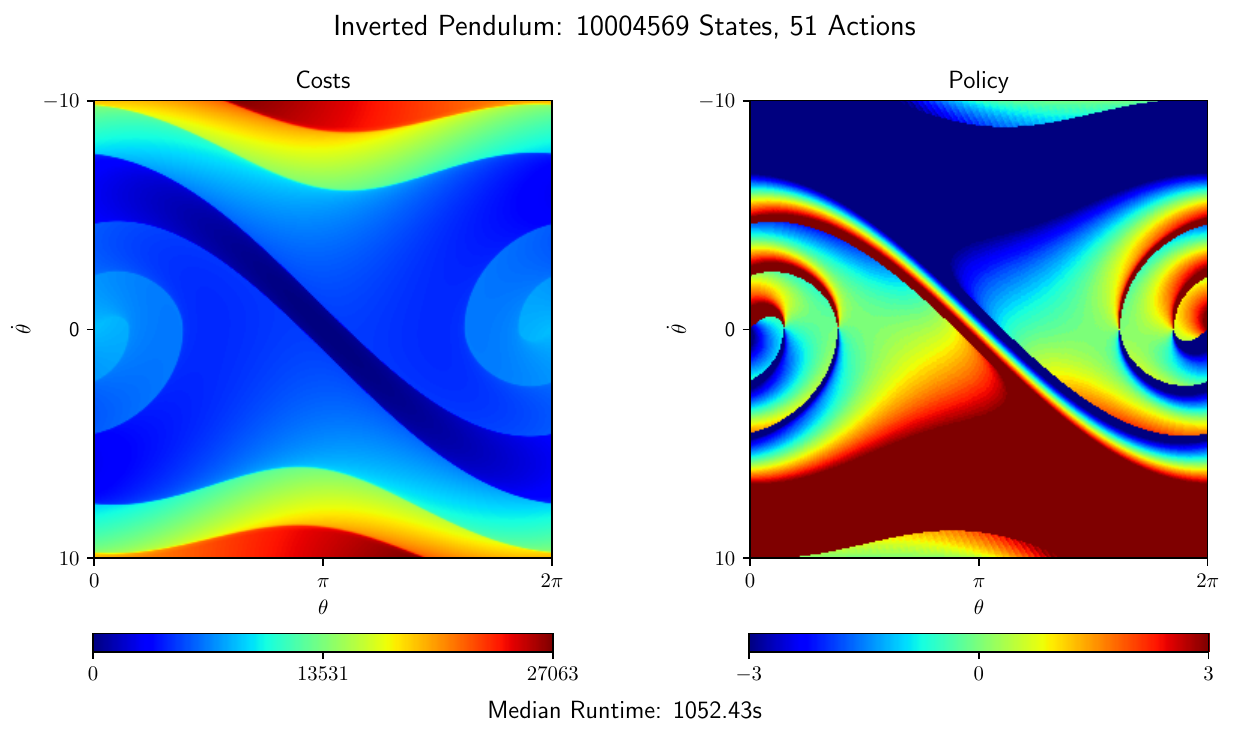}
   \caption{}
   \label{fig:high}
\end{subfigure}

\caption[]{Visualizations of the optimal cost (left) and optimal policy (right) computed by madupite for the discrete approximation of the inverted pendulum system using (a) $N_s = 11$ and (b) $N_s = 3163$ grid points.}
\label{fig:discretization}
\end{figure}

We can rewrite~\eqref{eq:inverted_pendulum_dynamic} using the following compact and generic form
\begin{equation}\label{eq:discretized_inverted_pend_eq}
    \dot{s}(\tau) = f(s(\tau), a(\tau))
\end{equation}
where $s(\tau) = \begin{bmatrix} \theta(\tau) \\ \dot{\theta}(\tau)
\end{bmatrix}$ and $a(\tau) = F(\tau)$.
After a time-discretization of~\eqref{eq:discretized_inverted_pend_eq} with forward Euler using $T_s = 0.01$ as time-step, we obtain
\begin{equation}\label{eq:time_discrete_ip}
    s_{t+1} = s_t + T_s f(s_t, a_t)\,.
\end{equation}
To abstract~\eqref{eq:time_discrete_ip} to an MDP, we also need to discretize the state and action spaces. We select $[0, 2\pi] \times [-10, 10]$ and $[-3, 3]$ as ranges for $s_{t}$ and $a_t$, respectively. We then divide each range into intervals using a uniform grid. We discretize each dimension of the state using $N_s$ grid points (giving rise to a total of $N_s^2$ grid points) and the input using $N_a$ grid points. To investigate scalability, we set $N_a = 51$ and keep it constant across our experiments and we increase $N_s$ from 11 to 3163.  
 As illustrated in Figure~\ref{fig:discretization}, the higher granularity of discretization, the more accurate is the resulting approximation of the original continuous system in~\eqref{eq:time_discrete_ip}. However, as expected, a higher granularity soon requires distributed storage and computation, which is possible thanks to madupite's distributed storage and computation capabilities.  Finally, we select $\gamma = 0.999$ to approximate the undiscounted scenario. We set $\alpha=0.001$, $tol=1e-7$ and use TFQMR as inner solver.
 We run the benchmarks on Euler, the ETH scientific computing cluster, using 32 cores~\cite{euler_cluster}. As anticipated, Figure~\ref{fig:pendulum} illustrates that computational costs increase with the problem size. However, madupite successfully computes the optimal solution even for state spaces of the order of millions in less than 1 hour. Compared to the infectious disease case, the problem appears to be significantly more complex. This complexity accounts for the longer convergence times, primarily driven by the greater number of iterations required. iPI requires indeed around 130 iterations to converge to the desired tolerance, which is significantly higher than its usual average of fewer than 20 iterations.

\subsection{2D Maze}\label{subsec: maze}
We consider a deterministic 2D grid-world environment defined over a rectangular lattice of size $H \times W$. The environment consists of a finite set of states $\mathcal{S} = \{0, 1, \dots, HW - 1\}$, where each state $s$ corresponds to a unique grid cell indexed in row-major order. A subset of states, denoted $\mathcal{O} \subset \mathcal{S}$, represents static obstacles that are not directly encoded in the transition function but emerge implicitly through the dynamics: any state with no valid outgoing transitions is treated as an obstacle and excluded from reachable paths. At each time step, an agent in state $s$ selects an action $a \in \mathcal{A} = \{0, 1, 2, 3, 4\}$, corresponding to \emph{stay}, \emph{north}, \emph{east}, \emph{south}, and \emph{west}, respectively. The environment exhibits deterministic transition dynamics: each action deterministically maps to a unique successor state $s'$, unless the move would result in exiting the grid boundaries or entering a state identified as unreachable, \textit{i.e.}, an obstacle. In such cases, the agent remains in its current state, and a large penalty cost is incurred. The objective is to reach a designated goal state $s_g$, which corresponds to the bottom-right cell, from any initial location while minimizing the accumulated stage cost. Free cells are associated with zero cost by default, while transitions into or through wall cells incur a prohibitive cost of $+\infty$ (approximated with $1e20$), discouraging any feasible policy from passing through them. The goal state is located near the bottom-right of the maze and is assigned a strongly negative cost, typically $-100$, to incentivize the agent to reach it.

\begin{figure*}
    \centering
    \includegraphics[width=0.8\textwidth]{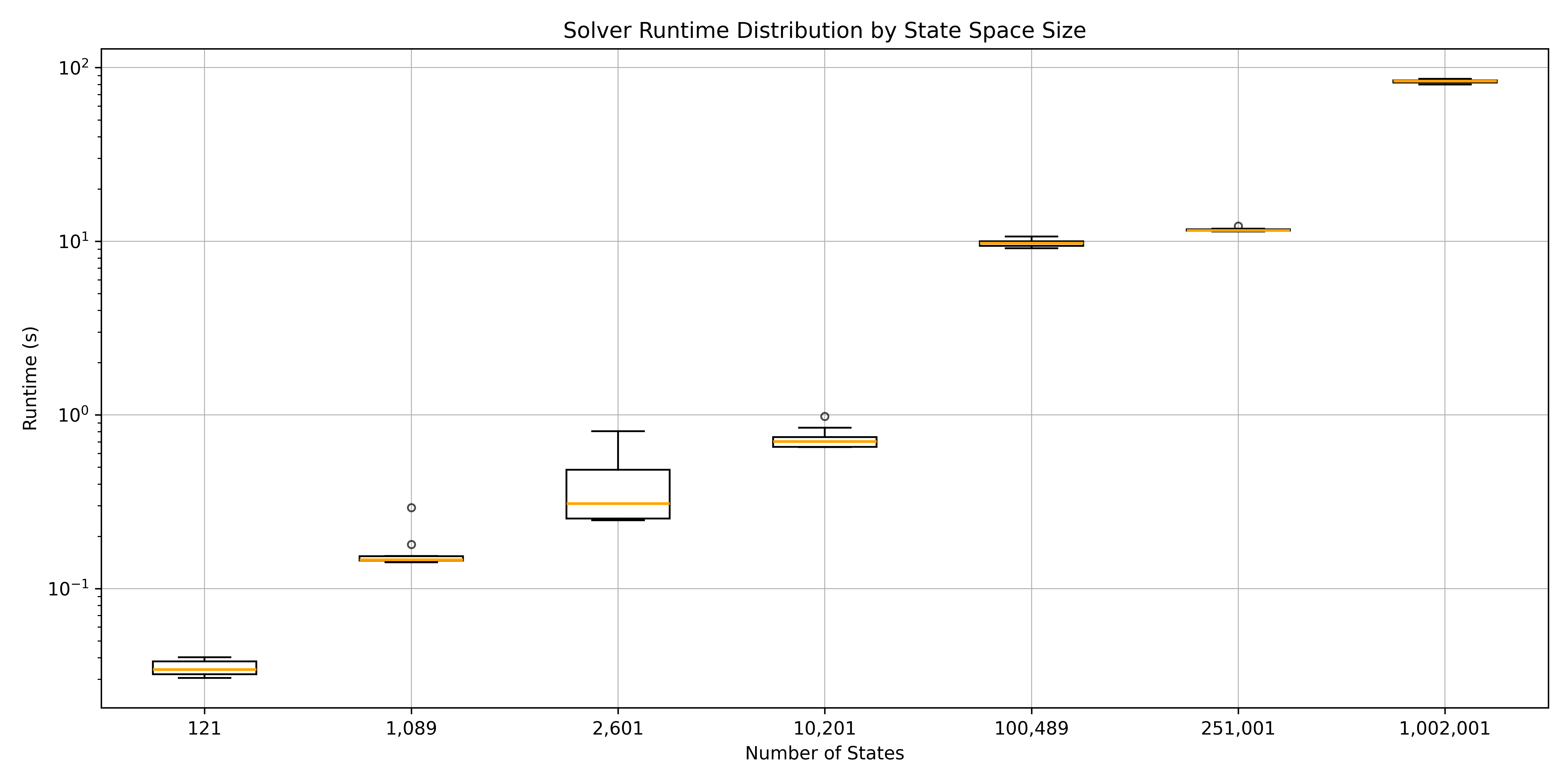}
    \caption{Runtime versus number of states when madupite is used to solve MDPs representing navigation tasks in 2D mazes of varying sizes. The benchmarks are run on 32 cores of the Euler cluster.}
    \label{fig:maze}
\end{figure*}

\begin{figure}
\vspace{-1cm}
    \centering
    \includegraphics[width=\linewidth]{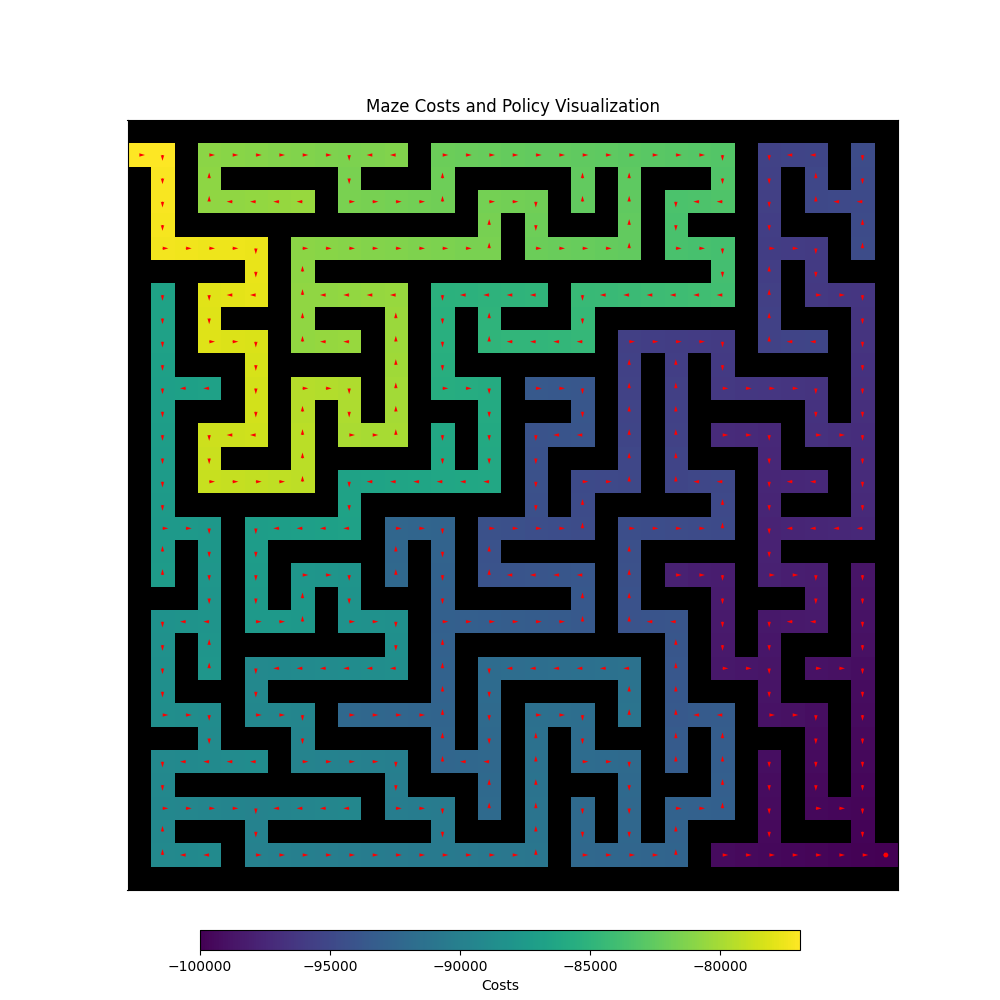}
    \caption{Graphical representation of a $33\times 33$-maze. The heatmap shows the optimal cost, and the red arrows indicate the optimal policy computed by madupite.}
    \label{fig:maze_visual}
\end{figure}

This formulation yields a large-scale sparse, deterministic planning problem with geometric constraints, well suited for madupite. We used $\gamma = 0.99$, TFQMR as the inner solver, and set $\alpha = 0.0001$. All benchmarks were executed on 32 cores of the Euler cluster~\cite{euler_cluster}. The plot in Figure~\ref{fig:maze} shows the average solver runtimes in seconds, computed over 10 independent runs. As shown in the plot, madupite can efficiently solve even the largest instance, comprising approximately 1 million states, in under 1.6 minutes. Once again, this performance is achieved through efficient distributed memory storage and computation, minimal communication overhead, and smart algorithm selection - all features which are unique to madupite. Finally, Figure~\ref{fig:maze_visual} provides a graphical representation of the maze with obstacles, along with the optimal policy (indicated by red arrows) and the optimal cost (shown as a heatmap) computed by madupite for $W = H = 33$.

\section{CONCLUSION}\label{sec: conclusions}
We presented the technical details and implementation choices behind madupite, and clarified the connection between its user-selectable parameters and the underlying mathematical algorithmic framework. We also thoroughly demonstrated its performance, comparing it against existing SOTA solvers to highlight its superior efficiency. Our evaluation includes standard metrics such as strong scaling, as well as case studies, both inherently discrete and obtained by discretizing a continuous MDP. Our numerical evaluation showcases that madupite establishes a new state-of-the-art by distributing both storage and computation, while leveraging more efficient algorithms that ensure fast convergence even in settings with large discount factors.

A promising future direction is extending the codebase to support risk-averse MDPs, as the solution methods proposed in~\cite{gargiani_riskaverse} are highly parallelizable and well-suited to madupite's framework.

\section*{ACKNOWLEDGMENT}
This work was supported by the European Research Council under the Horizon 2020 Advanced under Grant 787845 (OCAL).

\bibliographystyle{abbrv}
\bibliography{sn-article}

\end{document}